\documentclass[useAMS,usenatbib]{mn2e}

\usepackage[T1]{fontenc}
\usepackage{pslatex}
\usepackage{amsmath}
\usepackage{amsfonts}
\usepackage{color}
\usepackage{url}
\usepackage{graphicx}
\usepackage{subfigure}
\usepackage{amssymb}
\usepackage{textcomp}
\usepackage{soul}
\usepackage{bm}
\usepackage{booktabs}
\usepackage{array}
\usepackage{hyperref}
 \graphicspath{{images/}}

{\newif\ifnotend
\notendtrue
\def\veclist{ABCDEFGHIJKLMNOPQRSTUVWXYZabcdefghijklmnopqrstuvwxyz.}
\def\top#1#2.{#1}
\def\tail#1#2.{#2.}
\loop\expandafter\xdef\csname v\expandafter\top\veclist\endcsname%
{{\noexpand\bf\expandafter\top\veclist}}
\edef\veclist{\expandafter\tail\veclist}
\if\veclist.\notendfalse\fi\ifnotend\repeat}

\def\pa{\partial}

\mathchardef\mhyphen="2D

\title[Models of rotating coronae]{Models of rotating coronae}

\author[Sormani, Sobacchi, Pezzulli, Binney, Klessen]{Mattia C. Sormani$^{1}$, Emanuele Sobacchi$^{2,3}$, Gabriele Pezzulli$^{4}$, James Binney$^{5}$,
\newauthor Ralf S. Klessen$^{1,6}$\\
$^1$Universit\"{a}t Heidelberg, Zentrum f\"{u}r Astronomie, Institut f\"{u}r theoretische Astrophysik, Albert-Ueberle-Str. 2, 69120 Heidelberg, Germany \\
$^2$Physics Department, Ben-Gurion University, P.O.B. 653, Beer-Sheva 84105, Israel \\
$^3$Department of Natural Sciences, The Open University of Israel, 1 University Road, P.O.B. 808, Raanana 4353701, Israel \\
$^4$Department of Physics, ETH Zurich, Wolfgang-Pauli-Strasse 27, 8093 Zurich, Switzerland \\
$^5$Rudolf Peierls Centre for Theoretical Physics, 1 Keble Road, Oxford OX1 3NP, UK \\
$^6$Universit\"at Heidelberg, Interdiszipli\"ares Zentrum f\"ur Wissenschaftliches Rechnen, Im Neuenheimer Feld 205, 69120 Heidelberg, Germany
}

\begin{document}

\date{}

\def\p{\partial}
\def\Omegap{\Omega_{\rm p}}

\newcommand{\di}{\mathrm{d}}
\newcommand{\bfx}{\mathbf{x}}
\newcommand{\bfe}{\mathbf{e}}
\newcommand{\bfxi}{\bm{\xi}}
\newcommand{\vlos}{\mathrm{v}_{\rm los}}
\newcommand{\Tspin}{T_{\rm s}}
\newcommand{\Tb}{T_{\rm b}}
\newcommand{\degree}{\ensuremath{^\circ}}
\newcommand{\Th}{T_{\rm h}}
\newcommand{\Tc}{T_{\rm c}}
\newcommand{\bfr}{\mathbf{r}}
\newcommand{\bfv}{\mathbf{v}}
\newcommand{\bfu}{\mathbf{u}}
\newcommand{\pc}{\,{\rm pc}}
\newcommand{\kpc}{\,{\rm kpc}}
\newcommand{\Myr}{\,{\rm Myr}}
\newcommand{\Gyr}{\,{\rm Gyr}}
\newcommand{\kms}{\,{\rm km\, s^{-1}}}
\newcommand{\de}[2]{\frac{\partial #1}{\partial {#2}}}
\newcommand{\cs}{c_{\rm s}}
\newcommand{\rb}{r_{\rm b}}
\newcommand{\rqu}{r_{\rm q}}
\newcommand{\nuP}{\nu_{\rm P}}
\newcommand{\thetaobs}{\theta_{\rm obs}}
\newcommand{\hatn}{\hat{\textbf{n}}}
\newcommand{\hatt}{\hat{\textbf{t}}}
\newcommand{\hatx}{\hat{\textbf{x}}}
\newcommand{\haty}{\hat{\textbf{y}}}
\newcommand{\hatz}{\hat{\textbf{z}}}
\newcommand{\hatX}{\hat{\textbf{X}}}
\newcommand{\hatY}{\hat{\textbf{Y}}}
\newcommand{\hatZ}{\hat{\textbf{Z}}}
\newcommand{\hatN}{\hat{\textbf{N}}}
\newcommand{\hater}{\hat{\mathbf{e}}_r}
\newcommand{\hateR}{\hat{\mathbf{e}}_R}
\newcommand{\hatephi}{\hat{\mathbf{e}}_\phi}
\newcommand{\hatez}{\hat{\mathbf{e}}_z}
\newcommand{\hateP}{\hat{\mathbf{e}}_P}
\newcommand{\hatePhi}{\hat{\mathbf{e}}_\Phi}
\newcommand{\hatetheta}{\hat{\mathbf{e}}_\theta}
\newcommand{\hatemu}{\hat{\mathbf{e}}_\mu}
\newcommand{\hatenu}{\hat{\mathbf{e}}_\nu}
\newcommand{\hatePL}{\hat{\mathbf{e}}_{P\Lambda}}
\newcommand{\nablaPL}{\nabla_{P\Lambda}}

\maketitle

\begin{abstract}
Fitting equilibrium dynamical models to observational data is an essential step in understanding the structure of the gaseous hot haloes that surround our own and other galaxies. However, the two main categories of models that are used in the literature are poorly suited for this task: (i) simple barotropic models are analytic and can therefore be adjusted to match the observations, but are clearly unrealistic because the rotational velocity $v_\phi(R,z)$ does not depend on the distance $z$ from the galactic plane, while (ii) models obtained as a result of cosmological galaxy formation simulations are more realistic, but are impractical to fit to observations due to high computational cost. Here we bridge this gap by presenting a general method to construct axisymmetric baroclinic equilibrium models of rotating galactic coronae in arbitrary external potentials. We consider in particular a family of models whose equipressure surfaces in the $(R,z)$ plane are ellipses of varying axis ratio. These models are defined by two one-dimensional functions, the axial ratio of pressure $q_{\rm axis}(z)$ and the value of the pressure $P_{\rm axis}(z)$ along the galaxy's symmetry axis. These models can have a rotation speed $v_\phi(R,z)$ that realistically decreases as one moves away from the galactic plane, and can reproduce the angular momentum distribution found in cosmological simulations. The models are computationally cheap to construct and can thus be used in fitting algorithms. We provide a python code that given $q_{\rm axis}(z)$, $P_{\rm axis}(z)$ and $\Phi(R,z)$ returns $\rho(R,z)$, $T(R,z)$, $P(R,z)$, $v_\phi(R,z)$. We show a few examples of these models using the Milky Way as a case study.
\end{abstract}


\section{Introduction}

Since the suggestion of \cite{Spitzer1956}, the existence of hot gaseous haloes (or coronae) surrounding disc galaxies has been widely discussed \citep[e.g.][]{Putman+2012}. In the early days their existence was uncertain and usually conjectured on the basis of early models of galaxy formation \citep[][]{Binney1977,WhiteRees1978}, but there is now conclusive observational evidence for the existence of such coronae. 

The main and only direct observational evidence of galactic coronae comes from X-ray studies of emission and absorption lines of highly ionised species, both for the Galaxy \citep[e.g.][]{Yoshino2009, MillerBregman2013,MillerBregman2015,HodgesKluck+2016} and for external galaxies \citep[e.g.][]{OSullivan+2007,AndersonBregman2011,Bogdan+2013,Bogdan+2015,Walker+2015,Anderson+2016}. These observations have the potential to constrain the dynamics of the coronae in addition to their temperature and density profiles: for example, measuring the Doppler shifts of the O{\sc VII} absorption lines toward an ensemble of AGNs, \citet{HodgesKluck+2016} ruled out a stationary halo and suggested that the hot gas contains an amount of angular momentum comparable to that in the stellar disc of the Galaxy.

For the Galaxy, indirect evidence for the presence of a corona also comes from: (i) a remarkable depletion of gas in all dwarf galaxies within $R\simeq 270 \kpc$, which is naturally explained in terms of gas ablation as the dwarfs move through a hot corona \citep[][]{NicholsBlandHawthorn2011,Gatto+2013,Emerick+2016,TepperGarciaBlandHawthorn2018}; (ii) observed gas stripping and tadpole morphologies in the Magellanic System, which are similarly explained as caused by hydrodynamical interaction with the coronal gas \citep[e.g.][]{Salem+2015}. Note also that the fact that the Magellanic Stream (MS) contains gas but not stars \citep[e.g.][]{DonghiaFox2016} and the fact that the Stream is extremely head-tail asymmetric suggest that the MS is not a purely gravitational phenomenon \citep[e.g.][]{Putman+2011,For+2014}; (iii) the measured pressures of high-velocity clouds (HVC), which are consistent with pressure equilibrium with a surrounding hot medium \citep[][]{Stanimirovic+2002,Fox+2005}. The disc-corona interface is also probed by the dispersion measure of pulsars with known reliable distances \citep[][]{Gaensler+2008}, which measures the integrated free electron density out to the pulsar's distances, and by neutral-hydrogen 21cm emission data \citep{Marasco+2011,Marasco+2012}, which is present in significant amounts out to one or more kpc above the disc (note that at those heights the gas cannot be pressure supported in the vertical direction). However, due to the sparsity of observations, the properties of galactic coronae such as their mass content and extension remain largely uncertain.

Models of galactic coronae that are used in the literature for comparison with observations fall into two main categories:\footnote{A notable exception are the analytic baroclinic models of \cite{Barnabe+2006}.} (i) simple analytic models, which are either spherical and non-rotating \citep[e.g.][]{Fang+2013,TepperGarcia+2015,QuBregman2018} or rotating on cylinders, so that the rotational velocity $v_\phi$ does not depend on the distance $z$ from the Galactic plane \citep[e.g.][]{HodgesKluck+2016,LiBregman2017,Pezzulli+2017}, and (ii) those obtained as a result of cosmological simulations \citep[e.g.][]{Crain+2010,VanDeVoortSchaye2012,Stinson+2012,Shen+2013,Ford+2013,Velliscig+2015,Bogdan+2015,VanDeVoort+2016,Correa+2018,Oppenheimer2018}. Models of type (i) have the advantage that their parameters can be adjusted to match observations, but are clearly not realistic because we know that hot haloes rotate and that their rotation velocity must decrease with height $z$ above the galactic plane, while models of type (ii) are more realistic but cannot easily be fitted to observations, because a search in a large parameter space using simulations would be too computationally expensive. In the literature there is therefore a gap between realistic models and models that can be fitted to observations.

It is therefore important to construct more realistic analytic models which allow for an arbitrary rotation $v_\phi(R,z)$ (which can decrease with height), and that are easy to construct and to compare with observations. In this paper, we develop a simple method that allows to construct general axisymmetric equilibria in a given external potential. The key advantage is that the method is computationally cheap and makes it easy to obtain $\rho(R,z)$, $T(R,z)$, $P(R,z)$, $v_\phi(R,z)$ and similar quantities, which can then be fed to fitting algorithms. We discuss in particular a family of models whose equipressure surfaces are ellipses, and provide an illustrative python script that constructs these models and returns the above quantities.\footnote{The code is publicly available at the GitHub repository {\sc COROPY} \url{https://github.com/sormani/coropy}}

The paper is structured as follows. In Sect. \ref{sec:characterise} we write down the basic equations. In Sect. \ref{sec:models} and \ref{sec:mwmodels} we describe a family of models whose equipressure surfaces are ellipses, and show some applications to the Milky Way. In Sect. \ref{sec:conclusion} we sum up and indicate directions for future work.

\section{Characterisation of rotating equilibria} \label{sec:characterise}

We now prove that rotating axisymmetric baroclinic\footnote{The word \emph{baroclinic} is used here to indicate that $P$ is a function of both $T$ and $\rho$, in contrast to \emph{barotropic} which indicates that $P$ depends only on $\rho$.} equilibria in an external potential $\Phi$ with arbitrary entropy and angular momentum distributions are completely characterised by their pressure distribution $P(R,z)$. In particular: (i) given $P(R,z)$ a baroclinic equilibrium is uniquely identified and it is possible to find it constructively, and viceversa (ii) given a baroclinic equilibrium, $P(R,z)$ is uniquely determined. Statement (ii) is trivial, so we only need to prove (i). 

The Euler equation for an axisymmetric rotating baroclinic equilibrium in an external potential $\Phi$ reduces to:
\begin{equation} \label{eq:steady1}
- \frac{v_\phi^2}{R} \hateR  = -\frac{\nabla P}{\rho} - \nabla \Phi ,
\end{equation}
where $P(R,z)$ is the pressure, $\rho(R,z)$ is the density, $\bfv = v_\phi \hatephi$ is the velocity, and $(R,z,\phi)$ denote standard cylindrical coordinates. The continuity equation is automatically satisfied, so the only requirement for an equilibrium to be valid is that it satisfies equation \eqref{eq:steady1}.

Let us assume that we are given the function $P=P(R,z)$, i.e. we are given the value of the pressure everywhere. We define the unit vector normal to the surfaces of constant pressure as
\begin{equation}
\hateP = \frac{\nabla P}{| \nabla P|} = \cos(\theta_P) \hateR + \sin(\theta_P) \hatez ,
\end{equation}
and the unit vector perpendicular to it as
\begin{equation} \label{eq:hatenu1}
\hatenu = \hatephi \times \hateP = \sin(\theta_P) \hateR - \cos(\theta_P) \hatez \;.
\end{equation}
Let us write the gravitational potential as
\begin{equation}
\nabla \Phi = g(R,z) \hatePhi ,
\end{equation}
where 
\begin{equation} \label{eq:hatePhi}
\hatePhi = \cos(\theta_\Phi) \hateR + \sin(\theta_\Phi) \hatez \;.
\end{equation}
Taking the dot product of Eq. \eqref{eq:steady1} with $\hatenu$ we obtain
\begin{equation} \label{eq:steady2}
\boxed{ v_\phi^2 = Rg\cos(\theta_\Phi)\left[1-\frac{\tan(\theta_\Phi)}{\tan(\theta_P)}\right] = R\left[\frac{\pa\Phi}{\pa R}-\frac{\pa P/\pa R}{\pa P/\pa z}\frac{\pa\Phi}{\pa z}\right] }
\end{equation}
This quantity is easily calculated if we know $\Phi(R,z)$ and $P(R,z)$. Note that $v_\phi$ {\it only depends on the shape of the surfaces of constant pressure, and not on the value that the pressure assumes on them}. Viceversa, if we know $\Phi$ and $v_\phi$ everywhere then we can recover the shape of the equipressure surfaces.

In order to have $v_\phi^2>0$, the shape of the surfaces of constant pressure needs to satisfy
\begin{equation}
\label{eq:condition}
\frac{\tan(\theta_\Phi)}{\tan(\theta_P)} < 1\;.
\end{equation}
This condition is that the surfaces of constant pressure must be everywhere ``flatter'' than the surfaces of constant potential: for example, if the potential is spherical then surfaces of constant pressure that are ellipses elongated along $R$ are allowed, while ellipses elongated along $z$ are not allowed. Finally, note that $v_\phi^2$ vanishes if $\theta_\Phi=\theta_P$, namely if $\nabla P$ and $\nabla\Phi$ are parallel.

Now taking the dot product of Eq. \eqref{eq:steady1} with $\hatez$, or equivalently taking the dot product with $\hateP$ and then using \eqref{eq:steady2}, we obtain:
\begin{equation} \label{eq:steady3}
\boxed{ \rho = -\frac{|\nabla P|}{g} \frac{\sin(\theta_P)}{\sin(\theta_\Phi)} = -\frac{\pa P/\pa z}{\pa\Phi/\pa z}}
\end{equation}
Thus we see that, given $P(R,z)$, equations \eqref{eq:steady2} and \eqref{eq:steady3} allow to calculate $v_\phi(R,z)$ and $\rho(R,z)$, so that the equilibrium state is completely determined. This proves statement (i). 

This provides an easy method to construct rotating baroclinic equilibria: simply choose a function $P(R,z)$ (with the topology of surfaces of constant pressure that satisfies the constraint mentioned above), and calculate the rest. Moreover, it proves that there is a one-to-one correspondence between the `space of baroclinic equilibria' and the space of the functions $P(R,z)$ that satisfy the constraints described above.

\subsection{Calculation of the other quantities}

As discussed above, once $P(R,z)$ is given one can calculate $\rho(R,z)$ and $v_\phi(R,z)$ by using Eqs. \eqref{eq:steady2} and \eqref{eq:steady3}. One can then obtain all the other quantities, and in this section we provide all the definitions used in this paper for reference. The angular velocity is defined as
\begin{equation}
\Omega(R,z) = \frac{v_\phi}{R},
\end{equation}
and the specific angular momentum as
\begin{equation}
l(R,z) = R v_\phi.
\end{equation}
We assume that the gas is described by an ideal equation of state (note that we did not have to assume an equation of state until now),
\begin{equation} \label{eq:eos}
P =n k T,
\end{equation}
where $T$ is the temperature, $k$ is the Boltzmann constant, $n = \rho / (\mu m_{\rm p} )$ is the number density of particles, $\mu$ is the mean molecular weight and $m_{\rm p}$ is the proton mass. In this paper, we adopt $\mu = 0.58$. The entropy is defined as
\begin{equation} \label{eq:entropy}
\sigma = \log (P \rho^{-\gamma}),
\end{equation}
where $\gamma$ is the adiabatic index and $\log$ indicates the natural logarithm. We adopt $\gamma = 5/3$, the value for monoatomic ideal gases. Note that $\sigma$ is dimensionless and a change of units simply amounts to the addition of an unimportant additive constant.\footnote{The values displayed in the plots below are calculated assuming units of $M_\odot^{1-\gamma} (100 \kms)^2 \kpc^{3(\gamma-1)}$.}

%\subsection{A word on notation}
%
%In this paper, we will use the subscript `axis' to denote quantities along the $z$ axis, i.e. for any given function $f(R,z)$ we define
%%

\section{Models with elliptical equipressure surfaces} \label{sec:models}
In Sect. \ref{sec:characterise} we have seen that the function $P(R,z)$ completely characterises baroclinic equilibria, and thus by varying this function one can in principle obtain all possible baroclinic equilibrium models. However, since $v_\phi(R,z)$ only depends on the shape of the surfaces of constant pressure and not on the value that the pressure assumes on them, it is convenient to split the construction of an equilibrium into two steps:
\begin{enumerate}
\item Prescribe the {\it shape} of the surfaces of constant pressure.
\item Prescribe the {\it value} of $P$ on the surfaces.
\end{enumerate}
During the first step one can adjust the surfaces to obtain the desired $v_\phi(R,z)$. Then the second step will determine the mass and temperature distributions of the corona.

In the following, we consider models whose equipressure surfaces are ellipses in the plane $(R,z)$. An ellipse is defined by
\begin{equation} \label{eq:obl3}
\frac{R^2}{a(\mu)^2} + \frac{z^2}{b(\mu)^2} = 1,
\end{equation}
where $\mu$ is a parameter that labels the ellipses and $q=b/a$ defines their axis ratio.\footnote{Note that for a spherical potential with elliptical equipressure surfaces, as we will consider in Sect. \ref{sec:mwmodels}, we have $\tan(\theta_\Phi)/\tan(\theta_P)=q^2$. Eq. \eqref{eq:steady2} can be therefore rewritten as $\left(v_{\rm \phi}/v_{\rm c}\right)^2 = 1-q^2$, where $v_{\rm c}^2 = R \pa \Phi/\di R = Rg\cos(\theta_\Phi)$ is the local circular velocity of the potential. Hence in this case the surfaces of constant ${v_{\phi}}/{v_{\rm c}}$ and the surfaces of constant $q$, which are the equipressure surfaces, coincide.} In this paper, we will use the subscript `axis' to denote quantities along the $z$ axis, i.e. for any given function $f(R,z)$ we define 
\begin{equation}
f_{\rm axis}(z) \equiv f(R=0,z).
\end{equation}
The distribution $P(R,z)$ and hence the elliptical models are then completely determined by the following two functions:
\begin{enumerate}
\item $q_{\rm axis}(z)$: the value of the axial ratio of pressure along the axis $(R=0,z)$;
\item $P_{\rm axis}(z)$: the value of the pressure along the axis $(R=0,z)$.
\end{enumerate}
Once these two quantities are specified, one can calculate $P(R,z)$ and hence $v_\phi(R,z)$, $\rho(R,z)$, $T(R,z)$, etc using the equations of Sect. \ref{sec:characterise}. In the next section we explore some explicit models by using the Milky Way as a case study. 

\section{Illustrative application to the Milky Way} \label{sec:mwmodels}

In this Section we explore some illustrative models which are tuned to reproduce some basic properties of the Milky Way. We start with an unrealistic model 1, and step by step we adjust it to make more realistic as we go on with the numbering. Table \ref{table:1} provides a summary of the models.

\begin{table*}
\centering
\begin{tabular}{llllrrr}
\hline
Name    		& $q_{\rm axis}$				&  $P_{\rm axis}$			& $T_{\rm axis}$         		& $M_{200,\rm cor}/M_\odot$ 	& $L_{200,\rm cor}/L_0$	& $\lambda$	\\
\hline
Model 1		& 1 (spherical)					& Eq. \eqref{eq:Paxis2}		& Isothermal		  		& $3.4\times10^{10}$ 	& $0$					& 0 			\\
Model 2		& Eqs. \eqref{eq:q1a}-\eqref{eq:q1b}	& Eq. \eqref{eq:Paxis2}		& Isothermal	  			& $4.0\times10^{10}$ 	& $0.91$					& 0.038		\\
Model 3		& Eqs. \eqref{eq:q2a}-\eqref{eq:q2b}	& Eq. \eqref{eq:Paxis2}		& Isothermal	  			& $3.9\times10^{10}$ 	& $0.45$					& 0.019		\\
Model 4		& 1 (spherical)					& Eq. \eqref{eq:Paxis3}		& Polytropic $\Gamma=5/3$  	& $2.8\times10^{10}$ 	& $0$					& 0 			\\
Model 5		& Eqs. \eqref{eq:q1a}-\eqref{eq:q1b}	& Eq. \eqref{eq:Paxis3}		& Polytropic $\Gamma=5/3$	& $3.1\times10^{10}$ 	& $0.73$					& 0.039		\\
Model 6		& Eqs. \eqref{eq:q2a}-\eqref{eq:q2b}	& Eq. \eqref{eq:Paxis3}		& Polytropic $\Gamma=5/3$	& $3.1\times10^{10}$ 	& $0.38$					& 0.021		\\
\hline
\end{tabular}
\caption{Models discussed in this paper. $M_{200, \rm cor}$ and $L_{200, \rm cor}$ are the total mass and total angular momentum of the corona contained in the virial sphere of radius $r_{\rm 200}~=~237\kpc$. $L_0~=~10^{14}M_\odot \kms \kpc$ represents the order of magnitude of the total angular momentum contained in the Milky Way stellar disk \citep[e.g.][]{Peebles1969}. $\lambda = j_{200, \rm cor} / (\sqrt2 \, r_{200} v_{200})$ is the spin parameter according to the definition of \protect\cite{Bullock+2001}, where $j_{200, \rm cor} = L_{200,\rm cor}/ M_{200,\rm cor}$ is the averaged specific angular momentum of the corona.}
\label{table:1}
\end{table*}

\subsection{Potential}

In order to keep things simple and illustrative, we use in this paper a spherical NFW potential \citep[][]{NFW96}:
\begin{equation} \label{eq:nfwphi1}
\Phi(R,z) = - 4 \pi G \rho_0 r_0^2 \ \frac{\log\left( 1 + r/r_0\right)}{r/r_0},
\end{equation}
where
\begin{equation}
r = \sqrt{R^2 + z^2}.
\end{equation}
We use the following values: $r_0 = 20 \kpc$ and $\rho_0 = 0.01 \ M_\odot / {\rm pc^2}$. These values are appropriate for the Milky Way and are similar to the best fit values of \cite{McMillan2017}. The virial radius is $r_{200} = 237 \, \kpc$. This is defined as the radius of the sphere that has an average density 200 times the critical density $\rho_c = 3 H_0^2/(8 \pi G)$, where we have taken $H_0 = 73 \rm \kms \, Mpc^{-1}$ \citep[e.g.][]{FreedmanMadore2010}. The virial mass is $M_{200} = 1.64 \times 10^{12} \, M_\odot$ and the virial velocity is $v_{200} = \sqrt{G M_{200} / r_{200}} = 173\, \kms$.

There is in principle no difficulty in using flattened or more complicated numerically integrated potentials to produce further models. The only constraint is to ensure that $v_\phi^2>0$, which requires the isobaric surfaces to be ``flatter'' than the equipotential surfaces (see Eq. \ref{eq:condition}).

\subsection{Normalisation of the models} \label{sec:norm}

\begin{figure*}
\centering
\includegraphics[width=1.0\textwidth]{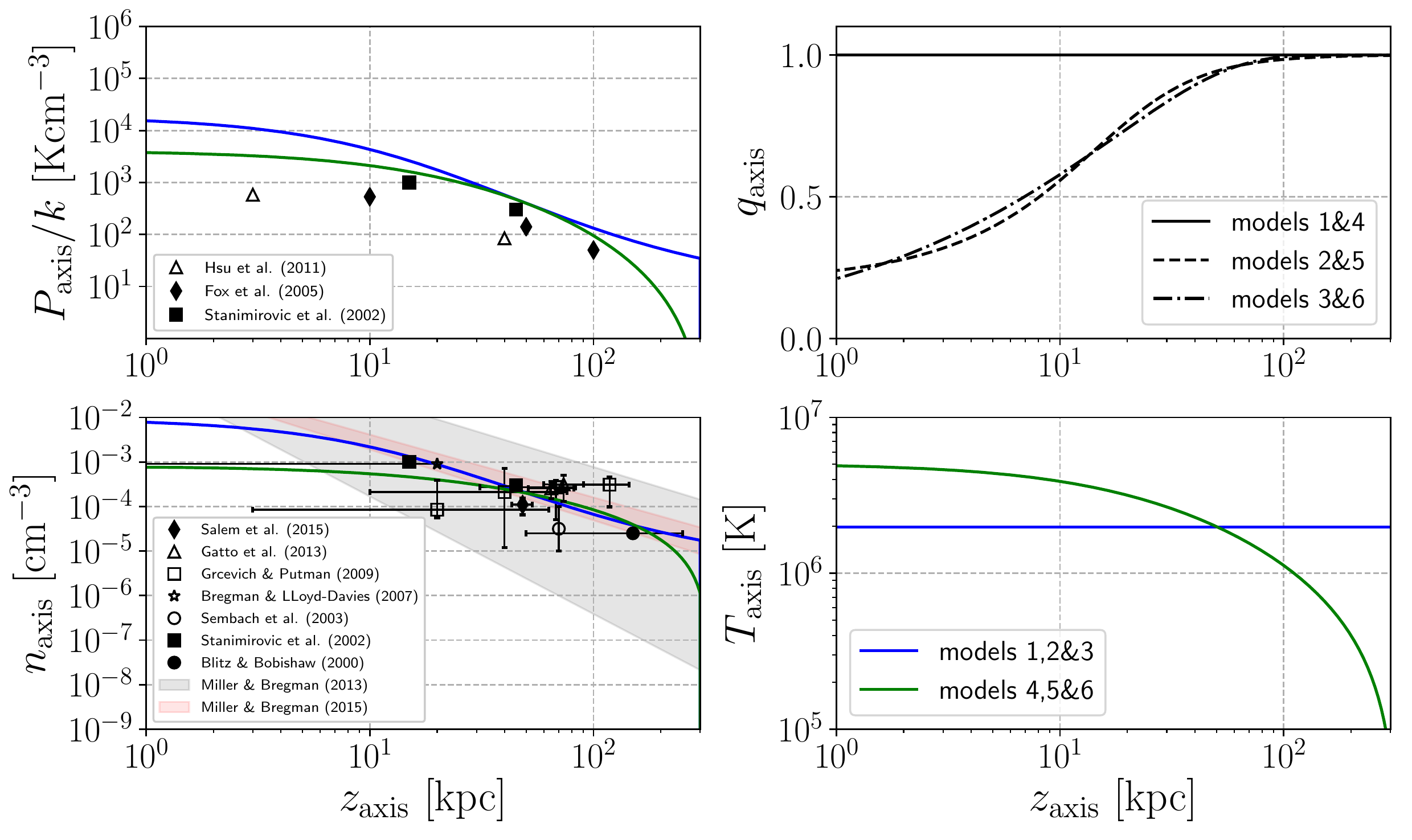}
\caption{Pressure, axis ratio, density and temperature profiles for the models discussed Sect. \ref{sec:mwmodels}. The scattered points represent estimates inferred from observations.}
\label{fig:panels}
\end{figure*}

The data points in Fig. \ref{fig:panels} show various estimates of density and pressure of the Milky Way corona at various distances inferred from observations (see table 7 of the review by \citealt{BlandHawthornGerhard2016}). The density estimates come from the following methods: (i) ram-pressure stripping arguments from satellite galaxies orbiting in the Galactic corona \citep{BlitzRobishaw2000,GrcevichPutman2009,Gatto+2013,Salem+2015} (ii) O{\sc VI} and O{\sc VII} absorption \citep{Sembach+2003,BregmanLLoydDavies2007,MillerBregman2013} (iii) O{\sc VIII} emission \citep{MillerBregman2015}. The pressure estimates all essentially come from estimating the pressure of warm ($T\gtrsim10^4 \ \rm K$) gas in High Velocity Clouds (HVCs), and then assuming that the hot corona is in pressure equilibrium with it \citep{Stanimirovic+2002,Fox+2005,Hsu+2011}.

Based on these measurements, we choose to normalise all our models so that $n_{\rm axis}= 2 \times 10^{-4} \ \rm cm^{-3}$ at $z = 50\kpc$. This approach is similar to that of \cite{TepperGarcia+2015} and, as also reported by them, it leads to a Galactic corona which broadly agrees with the results of observations of density over a broad range in distances. Interestingly, these models then all overestimate pressures. If instead one constructs models that match the observed pressures, density seem to be underestimated. Since measurements of pressure are all derived under the assumption of pressure equilibrium between the warm and hot medium, one possible interpretation is that the warm medium is at a slightly lower pressure than the hot medium. A similar conclusion was reached by \cite{Werk+2014} that, by analysing a sample of $L\sim L^{*}$ galaxies at redshift $z=0.2$, found that the pressure of the warm medium was substantially lower than needed to maintain pressure equilibrium with the hot medium. 

These considerations do not take into account that, since the spherical symmetry is broken in rotating coronae, one should also consider the full three dimensional geometry (i.e. the latitude and longitude of the various data points) when comparing models to observations. Huge uncertainties remain, and the challenge will be to construct a model which is consistent with as many observational constraints as possible simultaneously.

\subsection{Dispersion measures of pulsars} \label{sec:pulsar}

The red diamonds in Fig. \ref{fig:DMall} show the observed Dispersion Measures (DM) of pulsars with reliable distances. The DM is defined as
\begin{equation}
{\rm DM} = \int_0^d n_{\rm e}(l) \ \di l,
\end{equation}
where $n_{\rm e}(l)$ is the free electron density along the line of sight and $d$ is the distance to the pulsar. Since the main contribution to the observed DM is believed to come from the Warm Ionised Medium (WIM) in the disc \citep{Gaensler+2008},\footnote{Indeed, \citet{Howk+2006} compared a variety of ISM tracers, including the pulsar DM, in the foreground of the globular cluster NGC 5272 (Messier 3), which has $(l,b)=(42.2\degree,78.7\degree)$ and is located $z=10\kpc$ above the galactic plane. They found the warm ($T\sim 10^4\;$K) and hot ($T\gtrsim 10^5\;$K) ionised phases to be present in roughly a $5:1$ ratio along the line of sight.} which is not included in our models, one should not expect to fit these data with the coronal models alone. Instead, the observed DM provides an upper limit for the integrated free electron density in our coronal models. 

To calculate the DM in the models, we have assumed that the gas is completely ionised if $T\geq 10^4\ \rm K$, while it does not contribute if $T<10^4\ \rm K$, and that it is composed only of hydrogen and helium with proportions $75\%$ and $25\%$ in mass respectively as suggested by big-bang nucleosynthesis \citep[e.g.][]{Cyburt+2016}, so that $n_{\rm e} = 0.75 \times \rho/m_{\rm p} + 0.25 \times 2 \times \rho/(4 m_{\rm p})$ if $T\geq 10^4\ \rm K$. The position of the Sun is assumed to be at $(R_\odot,z_\odot) = (8\kpc, 0)$.

\subsection{Stability of the models}

Given an equilibrium, a natural question is whether it is dynamically stable or not. A useful check comes from the Solberg-H{\o}iland criteria, which state that a baroclinic equilibrium is dynamically stable with respect to isentropic axisymmetric motions if and only if the following two conditions are satisfied \citep[see for example][in particular his equations 3.94 and 3.95]{Tassoul2000}:
\begin{equation}
\frac{1}{R^3} \frac{\pa l^2}{\pa R} + \frac{1}{\gamma} \mathbf{g}_{\rm eff} \cdot \nabla \sigma > 0,
\end{equation}
\begin{equation}
g_{{\rm eff},z} \left( \frac{\pa l^2}{\pa R} \frac{\pa \sigma}{\pa z} - \frac{\pa l^2}{\pa z} \frac{\pa \sigma}{\pa R} \right) > 0,
\end{equation}
where
\begin{equation}
 \mathbf{g}_{\rm eff} = \left( \frac{\pa \Phi}{\pa z} \right)\hatez + \left(  \frac{\pa \Phi}{\pa R} - \frac{l^2}{R^3} \right) \hateR.
\end{equation}
We have numerically checked that for all the models discussed in the next subsection these criteria are satisfied.

\subsection{Models}

\begin{figure}
\centering
\includegraphics[width=0.5\textwidth]{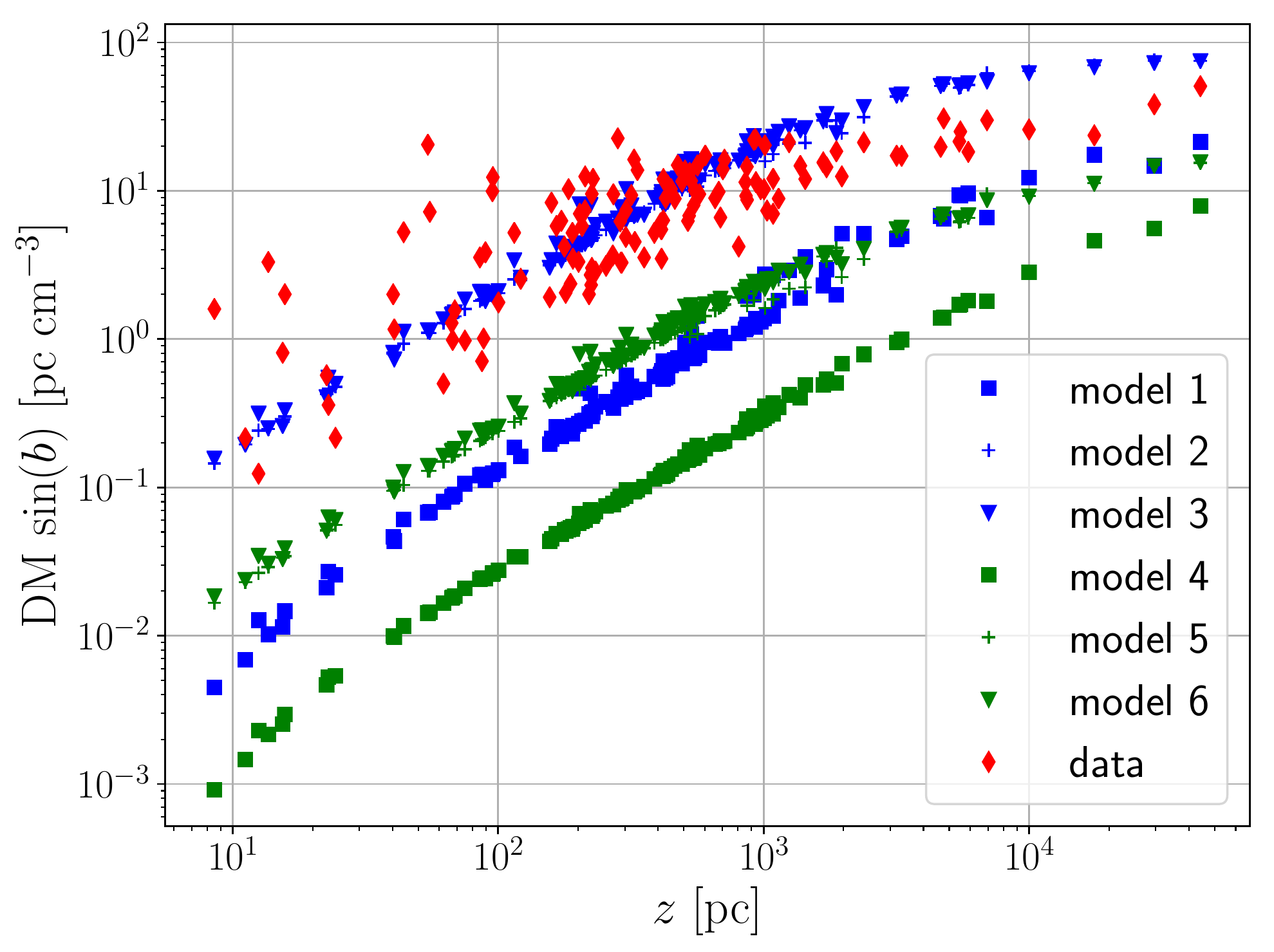}
\caption{Dispersion measures of pulsars with known reliable distances from observations (red diamonds) and calculated from our models. Following \protect\cite{Gaensler+2008}, we show here all the pulsars that fall in one of the following three categories: (i) pulsars in the ATNF Pulsar Catalogue \protect\citep[][available at \url{www.atnf.csiro.au/research/pulsar/psrcat}]{Manchester+2005} which have known parallaxes and DM; (ii) pulsars in globular clusters from the online compilation maintained by Paulo Freire at \url{http://www.naic.edu/~pfreire/GCpsr.html}. For each globular cluster, we plot only one point corresponding to the average DM of all the pulsars (which have all similar values for the same globular cluster), and use for the distance that from globular cluster read off the catalogue of globular clusters by \protect\cite{Harris1996} (2010 edition), \url{https://heasarc.gsfc.nasa.gov/W3Browse/all/globclust.html}; (iii) The two pulsars in the Magellanic Clouds listed by \protect\cite{Gaensler+2008}, with distances assumed to be $50\kpc$ and $61\kpc$ for the Large and Small Magellanic Cloud respectively.}
\label{fig:DMall}
\end{figure}

\subsubsection{Model 1}

We start with the simplest possible model, which will be useful for comparison with more complicate models later: a non-rotating, isothermal model. To build this model using the framework described in the previous sections, we need to find $q_{\rm axis}(z)$ and $P_{\rm axis}(z)$.

From Eq. \eqref{eq:steady2} we see that a model is non-rotating if and only if the equipressure and equipotential surfaces coincide. Since our potential \eqref{eq:nfwphi1} is spherical, the model will be non-rotating everywhere if and only if the equipressure surfaces are spheres. So for this model $q_{\rm axis}(z) = 1$. To find $P_{\rm axis}$, note that Eq. \eqref{eq:steady1} along the axis $(R=0,z)$ reduces to:
\begin{equation} \label{eq:Paxis1}
\rho_{\rm axis}(z) = -\frac{ P'_{\rm axis}}{\Phi'_{\rm axis}},
\end{equation}
where the superscript $'$ denotes derivative with respect to $z$. If we require the model to be isothermal along the $z$ axis (and thus by symmetry everywhere for this model), then $P_{\rm axis} = \cs^2 \rho_{\rm axis}$ where $\cs^2=k T/(\mu m_{\rm p})$ is a constant. Substituting this equation into \eqref{eq:Paxis1} and solving the differential equation we obtain:
\begin{equation} \label{eq:Paxis2}
P_{\rm axis} = P_0 \exp\left( - {\Phi_{\rm axis}}/{\cs^2} \right),
\end{equation}
where $P_0$ is a constant. We choose $\cs$ such that $T = 2 \times 10^6 \ \rm K$ and $P_0$ such that the normalisation of density is as described in Sect. \ref{sec:norm}.

Fig. \ref{fig:panels} shows the density and pressure profiles obtained for model 1. They are consistent with observations within the errors, although the model seems to overestimate the pressures as discussed in Sect. \ref{sec:norm}. Fig. \ref{fig:DMall} compares the observed Dispersion Measures (DM) of pulsars with known reliable distances (red diamonds) and the same quantities calculated in our models. As discussed in Sect. \ref{sec:pulsar}, our models should provide values well below the observed ones, because the main contribution should not come from the corona but from the warm ionised medium in the disc according to \cite{Gaensler+2008}. Model 1 is consistent with this expectation, although not by a large margin. However, we will see in the next section that when we make this model rotating (model 2) it will fail in this regard.

The main problem of model 1 is that it is not rotating. Hence, the next step is to make it rotate.

\subsubsection{Model 2}

\begin{figure}
\centering
\includegraphics[width=0.5\textwidth]{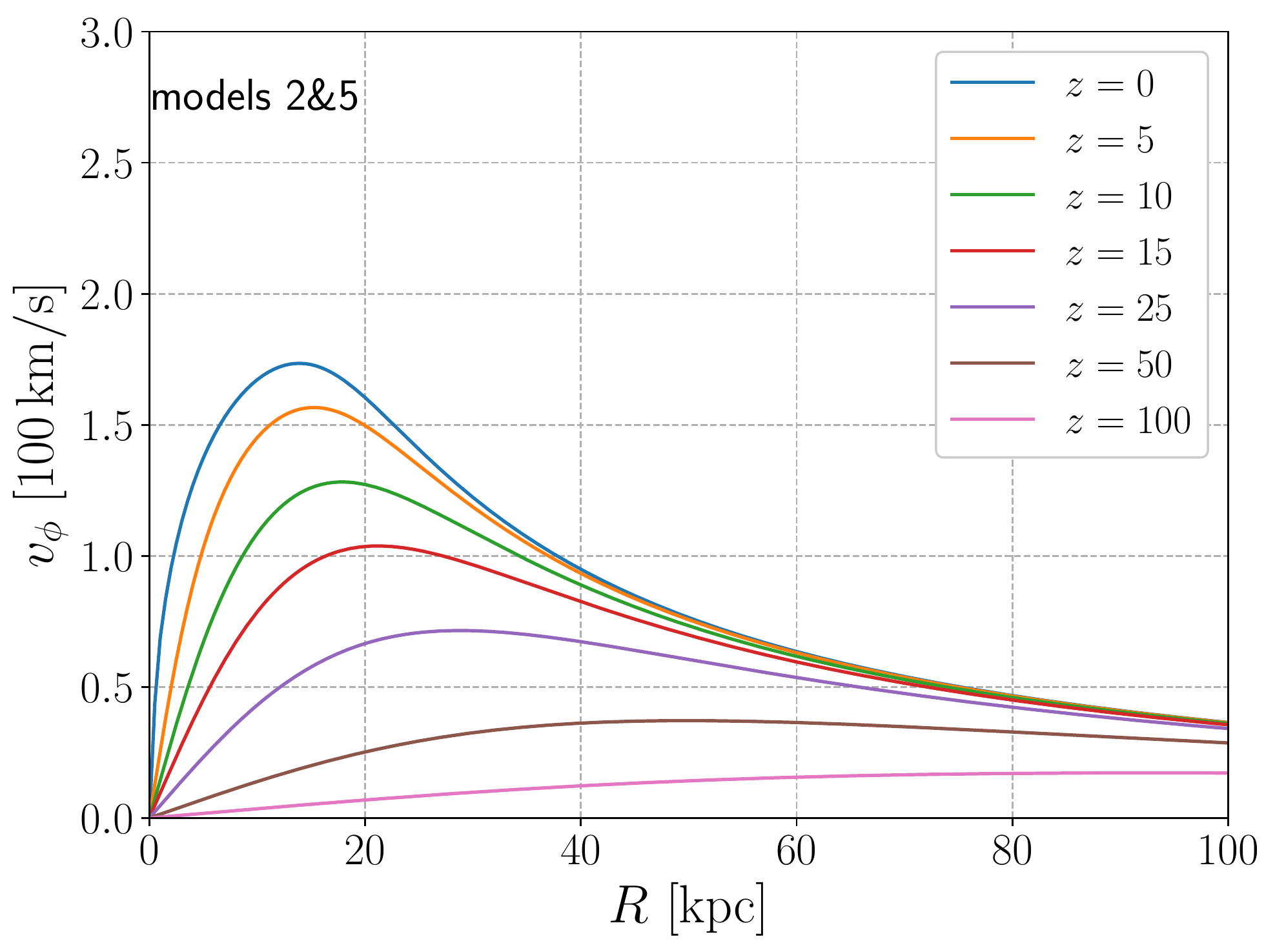}
\includegraphics[width=0.5\textwidth]{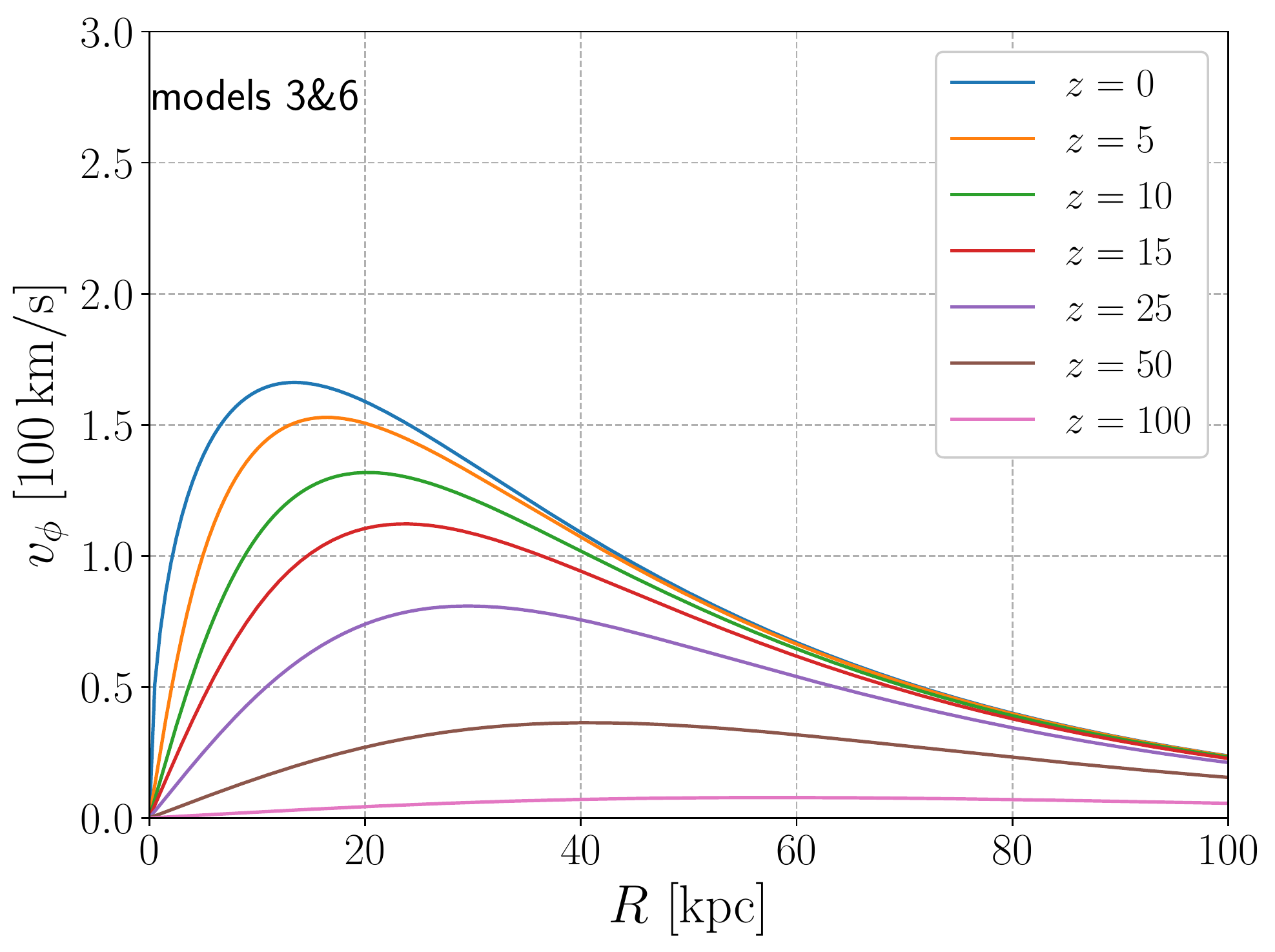}
\caption{Rotational velocity at a different heights from the Galactic plane. The top line is for $z=5\kpc$, and then at increasing heights at intervals of $5\kpc$. Top panel: model 2 and 5. Bottom panel: model 3 and 6.}
\label{fig:vphi}
\end{figure}

We want to modify model 1 to make it rotating. The minimal modification is to keep it isothermal along the $z$ axis, so we can take $P_{\rm axis}$ exactly as in model 1 (this works because Eq. \ref{eq:Paxis1} is unaffected by rotation). The rotation will make it not isothermal away from the axis.

What we need to change is $q_{\rm axis}$. We would like a model that rotates $\sim 80 \kms$ slower than the disc close to the plane, according to the findings of \cite{Marinacci+2011} and \cite{HodgesKluck+2016}, but reduces to the isothermal sphere of model 1 far away from the plane. To construct such a model we need equipressure surfaces that are elongated close to the plane but become spherical as we move away, i.e. $q_{\rm axis}<1$ close to the plane and $q_{\rm axis} \to 1$ as $r \to \infty$. A possible choice is:
\begin{align}
a(\mu) & = a_0 \frac{ \sinh(\mu)}{ [\eta + (1 - \eta) \tanh(\mu)] } \label{eq:q1a}, \\
b(\mu) & = b_0 \sinh(\mu) \label{eq:q1b}.
\end{align}
For $\eta=0$, this parametrisation reduces to confocal ellipses, i.e. the surfaces of constant pressure coincide with one of the coordinates in a \emph{oblate spheroidal coordinate} system. However, one can show from Eq. \eqref{eq:steady2} that all models with $\eta=0$ have the property that the rotational velocity close to the disc at $R<a_0$ tends to the circular velocity in the plane $z=0$, while we would like a corona that rotates roughly $\sim 80 \kms$ slower than the disc \citep{Marinacci+2011}. Moreover, the density and temperature become singular at the common focal point in these models. Choosing a positive value of $\eta$ solves both problems. For model 2, we choose $a_0=b_0=20\kpc$ and $\eta=0.2$.

The top-right panel in Fig. \ref{fig:panels} shows the resulting $q_{\rm axis}$. The top panel in Fig. \ref{fig:vphi} shows the rotational velocity at different heights above the plane. The rotational velocity is higher close to the plane and decreases going up. Fig. \ref{fig:m2} and \ref{fig:m2zoom} show various quantities in the $(R,z)$ plane. The contours of $v_\phi$ in Fig. \ref{fig:m2zoom} roughly follow the shapes obtained in cosmological simulations \citep[e.g.][]{Stinson+2010,Stinson+2012,Stinson+2013}. The temperature decreases close to the plane, hinting at a transition with a colder disc. Linear stability analysis usually conclude that coronae are stable to the thermal instability \citep{Binney+2009,Nipoti2010}, but assume that the gas is hot ($T\simeq10^6 \rm K)$. \cite{Binney+2009} find that thermal instability occurs if the coronal temperature falls through $3 \times 10^5 \ \rm K$, so it may be interesting to re-examine this issue using the current models, which close to the plane approach this temperature.

One problem of this model is that the DM of pulsars are too high. Making model 1 rotating has increased the DM dramatically. The reason is that the main contribution to the DM comes from regions close to the disc, and making the model rotating has made the density just above the Sun much higher (see bottom-right panel in Fig. \ref{fig:m2zoom} and compare with the spherical model 1). This is because now the disc is rotationally supported, and $n$ decreases much slower as a function of $R$ in the disc. This problem will be cured by increasing the temperature of the corona near the Galactic plane (models 4,5,6).

\begin{figure}
\centering
\includegraphics[width=0.5\textwidth]{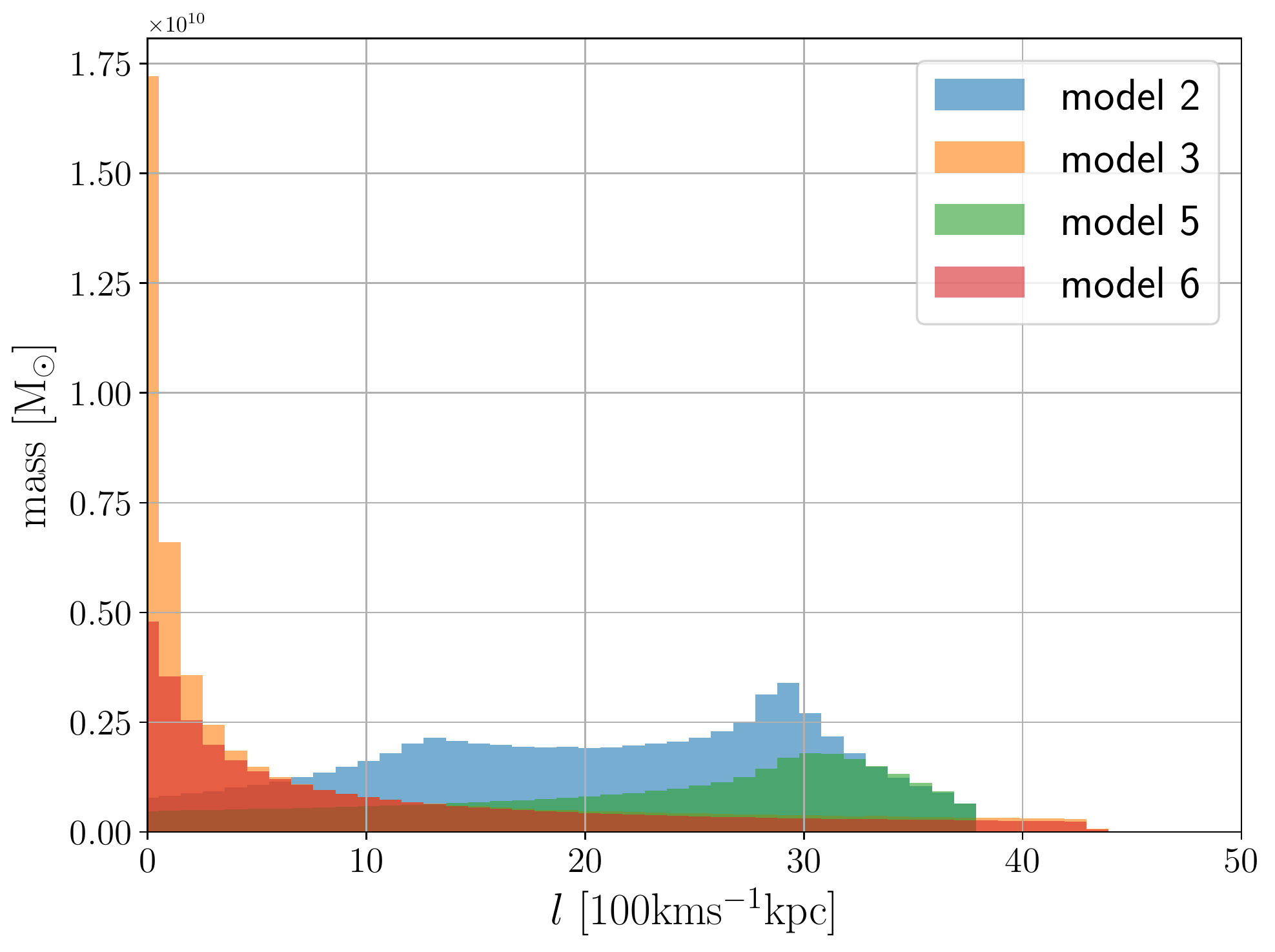}
\caption{Angular momentum distribution (AMD) for our models. The AMD is defined as the amount of mass in the corona per given angular momentum.}
\label{fig:AMD}
\end{figure}

\begin{figure}
\centering
\includegraphics[width=0.5\textwidth]{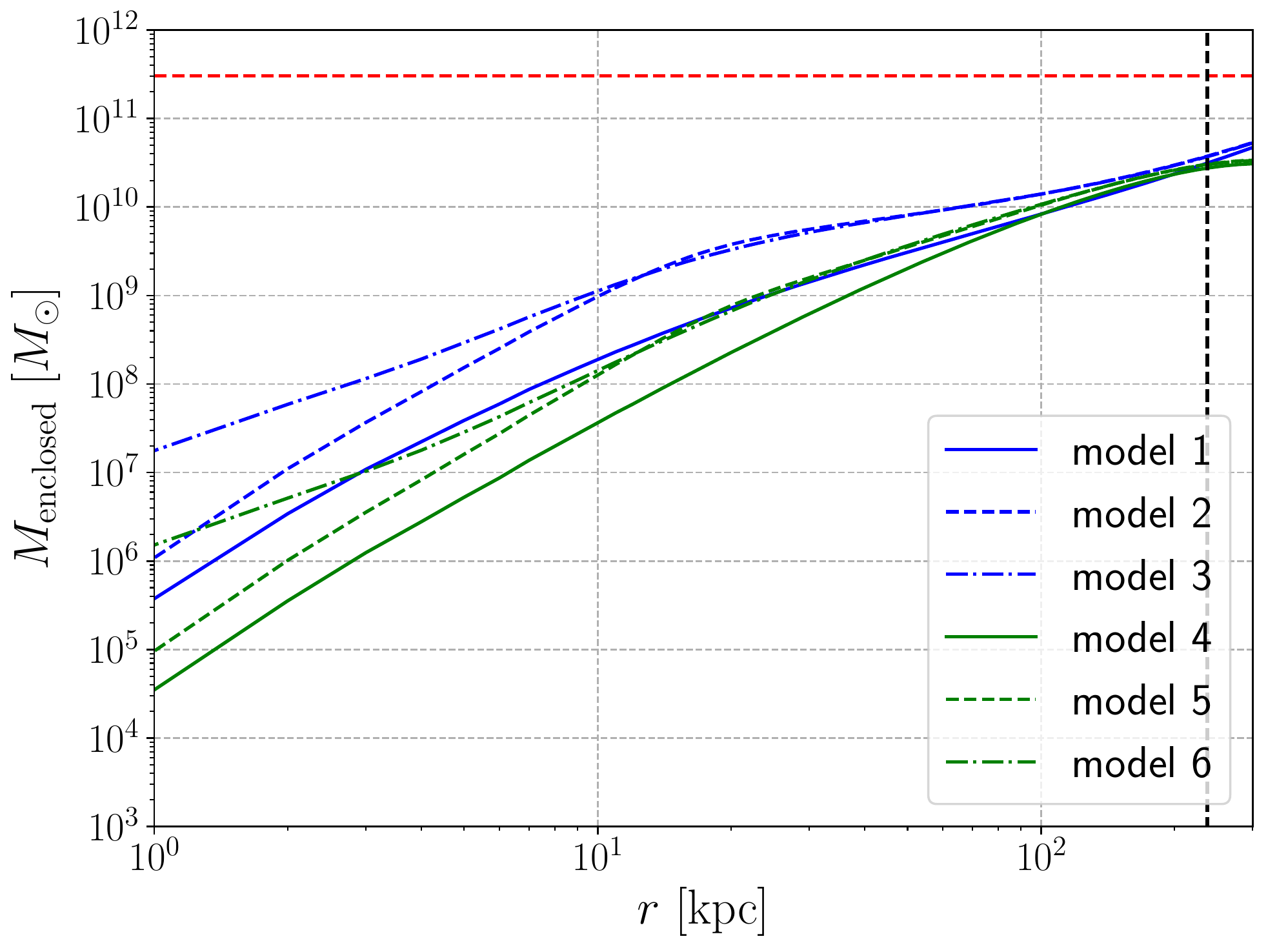}
\caption{Mass enclosed within spherical radius $r$ in our models. The black vertical dashed line indicates the virial radius $r_{200}$. The red horizontal line at $M_{200} \Omega_{\rm b}/\Omega_{\rm c} = 3 \times 10^{11} \, M_\odot$ indicates the baryons that should be contained within $r_{200}$ according to the cosmological value of the ratio of baryons to dark matter, where we have taken $\Omega_{\rm b}/\Omega_{\rm c}=18.6\%$ \citep{Planck2018}. }
\label{fig:Menc}
\end{figure}

Another problem of this model is shown by Fig. \ref{fig:AMD}, which shows the angular momentum distribution (AMD) for our models. The AMD is defined as the distribution of mass per unit angular momentum. Cosmological simulations typically find the AMD to be roughly exponential \citep[e.g.][]{VanDenBosch+2002,SharmaSteinmetz2005,Sharma+2012}, but that of model 2 is clearly not.\footnote{Since X-ray observations mostly probe the innermost $\lesssim50 \kpc$ of the corona, we have to rely on predictions from cosmological simulations to construct the outer parts ($R\gtrsim50\kpc$) of our models.} Since most of the mass is at outer radii (Fig. \ref{fig:Menc}), this indicates that there is an excess of rotation (angular momentum) at large radii.

To cure this problem, we need to modify the function $q_{\rm axis}$. This motivates model 3.

\subsubsection{Model 3}

\begin{figure*}
\centering
\includegraphics[width=1.0\textwidth]{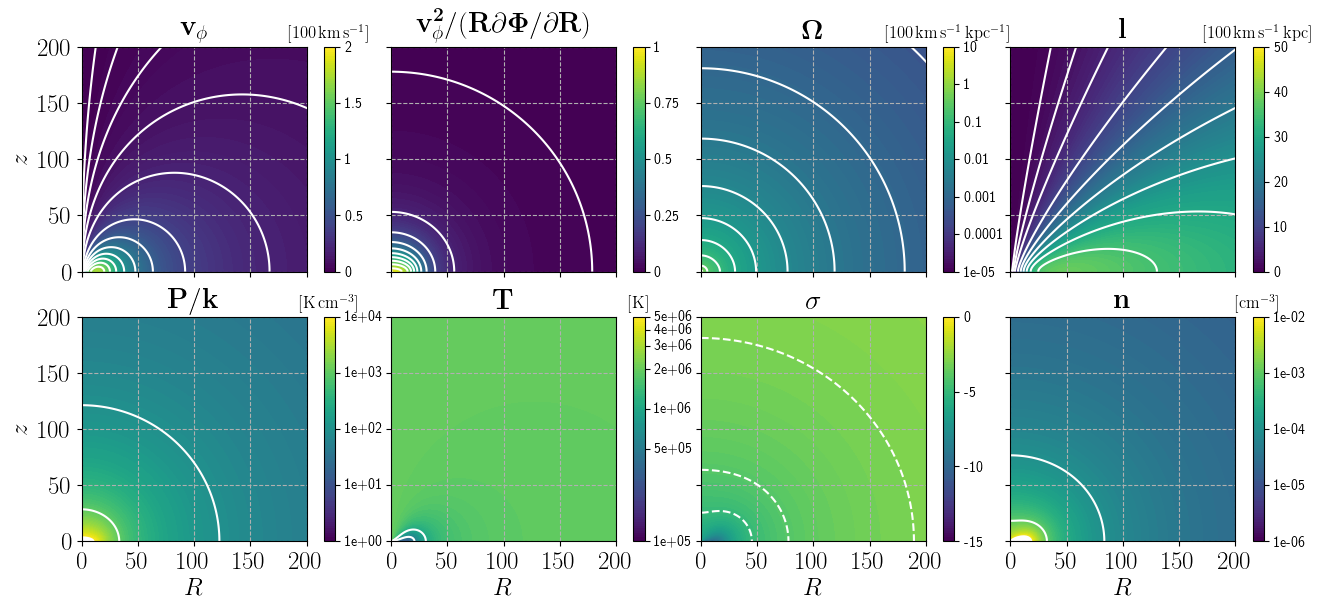}
\caption{Model 2. For $P/k$, $T$ and $n$, the white contours coincide with labels in the colorbars. The white dashed contours for $\sigma$ are at $\{-5,-4,-3\}$.}
\label{fig:m2}
\end{figure*}

\begin{figure*}
\centering
\includegraphics[width=1.0\textwidth]{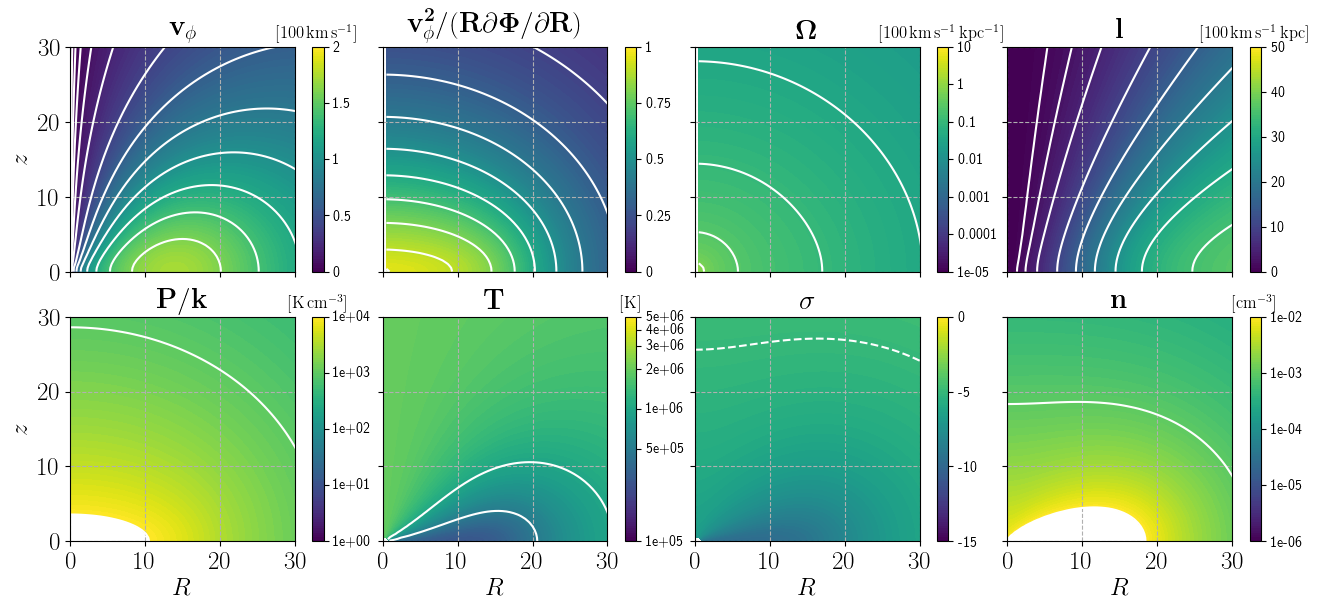}
\caption{Model 2. Zoom in the innermost 30 $\kpc$ of Fig. \ref{fig:m2}.}
\label{fig:m2zoom}
\end{figure*}

\begin{figure*}
\centering
\includegraphics[width=1.0\textwidth]{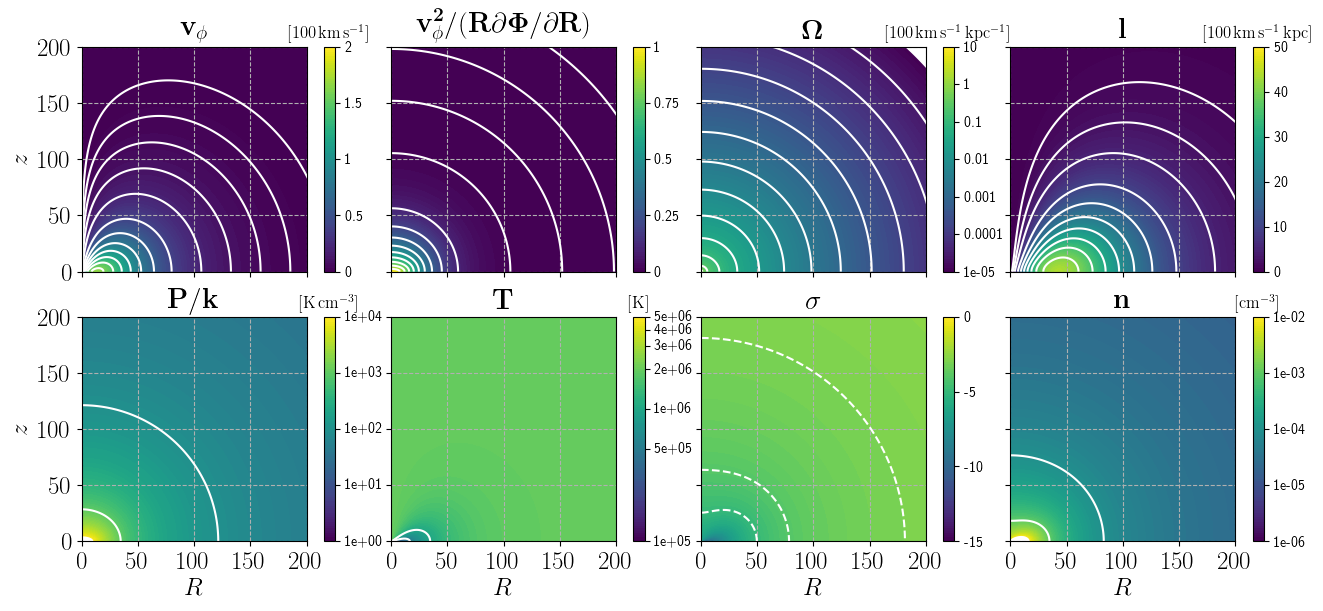}
\caption{Model 3. For $P/k$, $T$ and $n$, the white contours coincide with labels in the colorbars. The white dashed contours for $\sigma$ are at $\{-5,-4,-3\}$.}
\label{fig:m3}
\end{figure*}

\begin{figure*}
\centering
\includegraphics[width=1.0\textwidth]{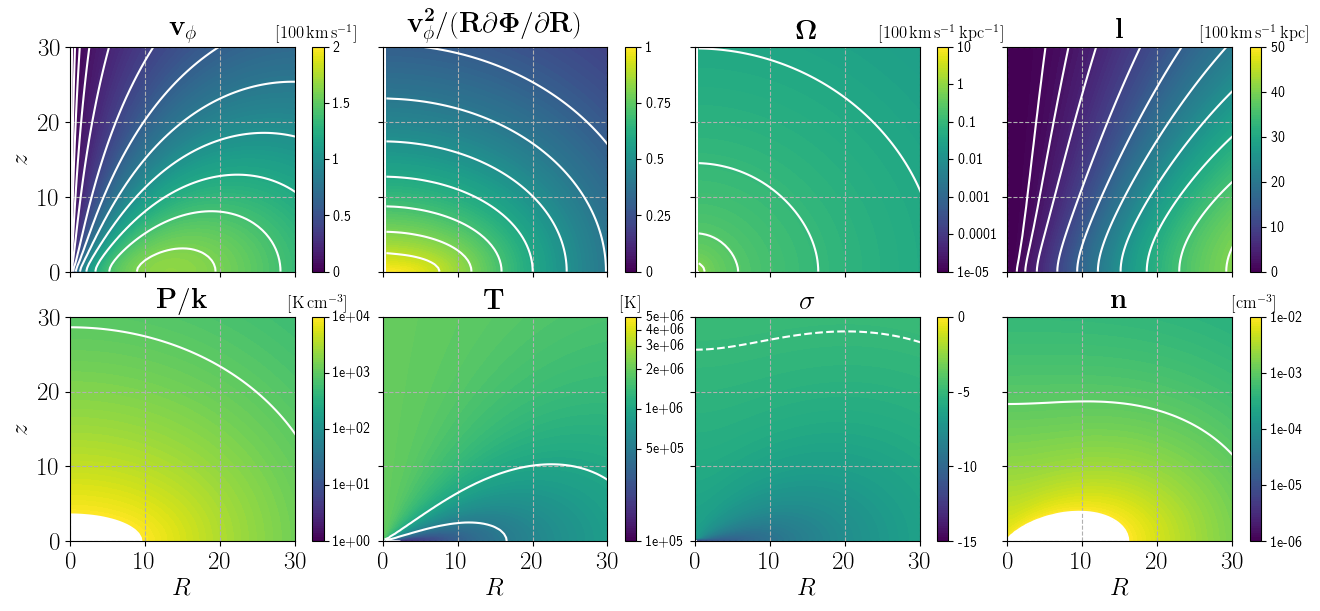}
\caption{Model 3. Zoom in the innermost 30 $\kpc$ of Fig. \ref{fig:m3}.}
\label{fig:m3zoom}
\end{figure*}

To cure the AMD problem encountered with model 2, we need to choose $q_{\rm axis}$ so that the corona rotates slower at large radii, where most of the mass is concentrated. Hence we consider the following parametrisation:
\begin{align}
a(\mu) & = \mu \label{eq:q2a} \\
b(\mu) & = \mu \left[ 1 - \exp(-\mu/L) \right] \label{eq:q2b}
\end{align}
The corresponding $q_{\rm axis}$ is shown in the top-right panel in Fig. \ref{fig:panels}. We have used $L = 20\kpc$. We see that model 3 rotates faster than model 2 for $R\lesssim 15\kpc$, but rotates slower for $R\gtrsim 15\kpc$. The difference is very subtle and is difficult to see by comparing the top and bottom panels in \ref{fig:vphi} or by comparing 2D maps as in Figs. \ref{fig:m2}, \ref{fig:m3} and \ref{fig:m2zoom}, \ref{fig:m3zoom}. Nevertheless, the difference in the AMD is quite large, and we see in Fig. \ref{fig:AMD} that the resulting AMD of model 3 is roughly exponential, as suggested by cosmological simulations.

This model retains the problem of model 2 that DM of pulsars is too high. In order to cure this problem, we need to rise the temperature of the corona close to the Galactic plane.

\subsubsection{Model 4}

In order to cure the problem with pulsars DMs of model 4, we need to find a model with higher temperature close to the Galactic plane. We start again from a spherical model, and instead of taking it isothermal, we take it polytropic, i.e. we assume that $P_{\rm axis}  \propto \rho_{\rm axis}^\Gamma$. We assume $\Gamma=5/3$. Substituting this into \eqref{eq:Paxis1} and solving the differential equation yields 
\begin{equation} \label{eq:Paxis3}
P_{\rm axis} = P_0 \left[ C - \Phi_{\rm axis} \right]^{\Gamma/(\Gamma-1)},
\end{equation}
where $C$ is a constant that controls the temperature profile and $P_0$ is a constant that controls the mass scaling. We choose these constants so that $T_{\rm axis}=2\times10^6\ \rm K$ at $z=50 \kpc$ and the density normalisation is as described in Sect. \ref{sec:norm}.

From the bottom-left panel in Fig. \ref{fig:panels}, we see that the density profile of this model at small radii is much shallower, hence the densities are much lower at small radii. This brings down the value of the DM, which was the problem of model 3. Now we need to make this model rotating.

\subsubsection{Model 5}

\begin{figure*}
\centering
\includegraphics[width=1.0\textwidth]{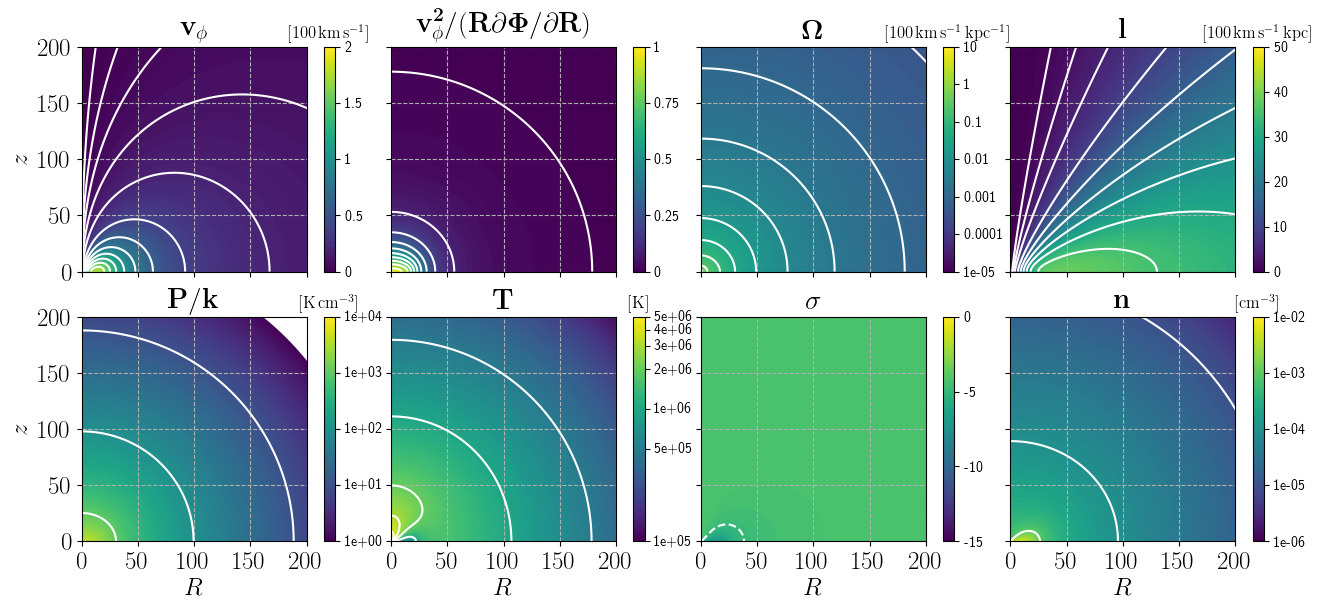}
\caption{Model 5. For $P/k$, $T$ and $n$, the white contours coincide with labels in the colorbars. The white dashed contours for $\sigma$ are at $\{-5,-4,-3\}$.}
\label{fig:m5}
\end{figure*}

\begin{figure*}
\centering
\includegraphics[width=1.0\textwidth]{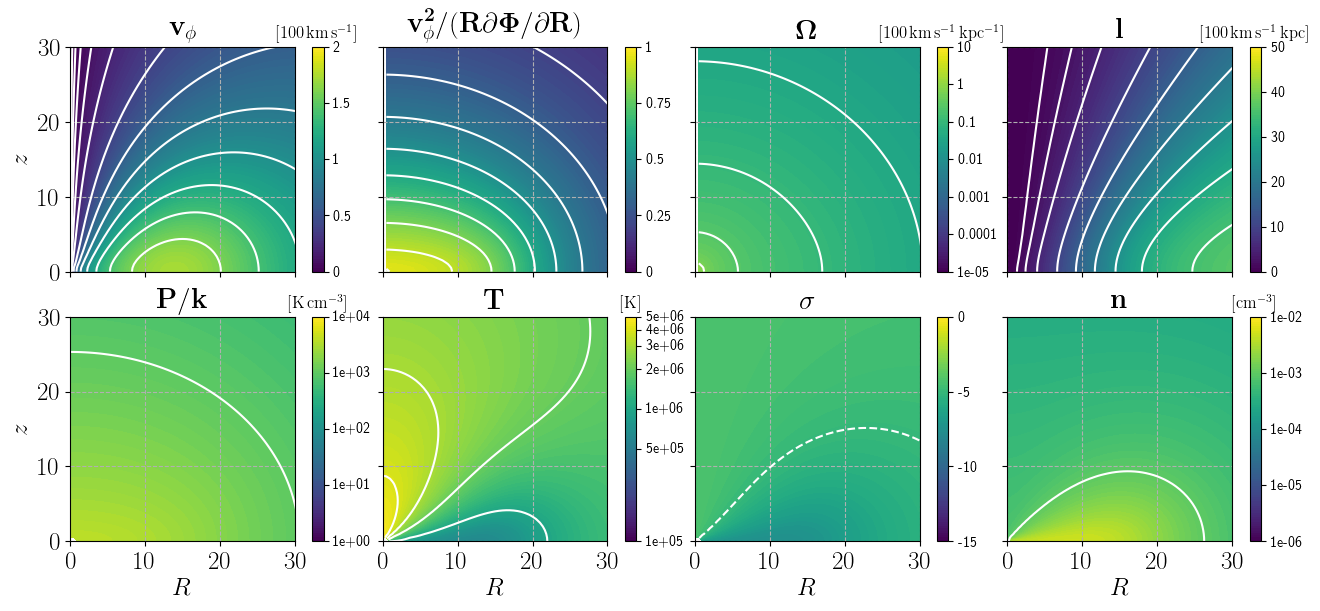}
\caption{Model 5. Zoom in the innermost 30 $\kpc$ of Fig. \ref{fig:m5}.}
\label{fig:m5zoom}
\end{figure*}

First we try to make model 4 rotating by modifying it in the same way we modified model 1 to obtain model 2. Thus for model 5 we keep the same $P_{\rm axis}$ as model 4, but we take $q_{\rm axis}$ as in model 2. The result is shown in Figs. \ref{fig:m5}, \ref{fig:m5zoom}. We see from Fig. \ref{fig:DMall} that this model solves the DM problem that plagued model 2 and 3, but we see from Fig. \ref{fig:AMD} that it still has the AMD problem that plagued model 2. To solve this, we can make the same modification to $q_{\rm axis}$ that we made in going from model 2 to model 3.

\subsubsection{Model 6}

\begin{figure*}
\centering
\includegraphics[width=1.0\textwidth]{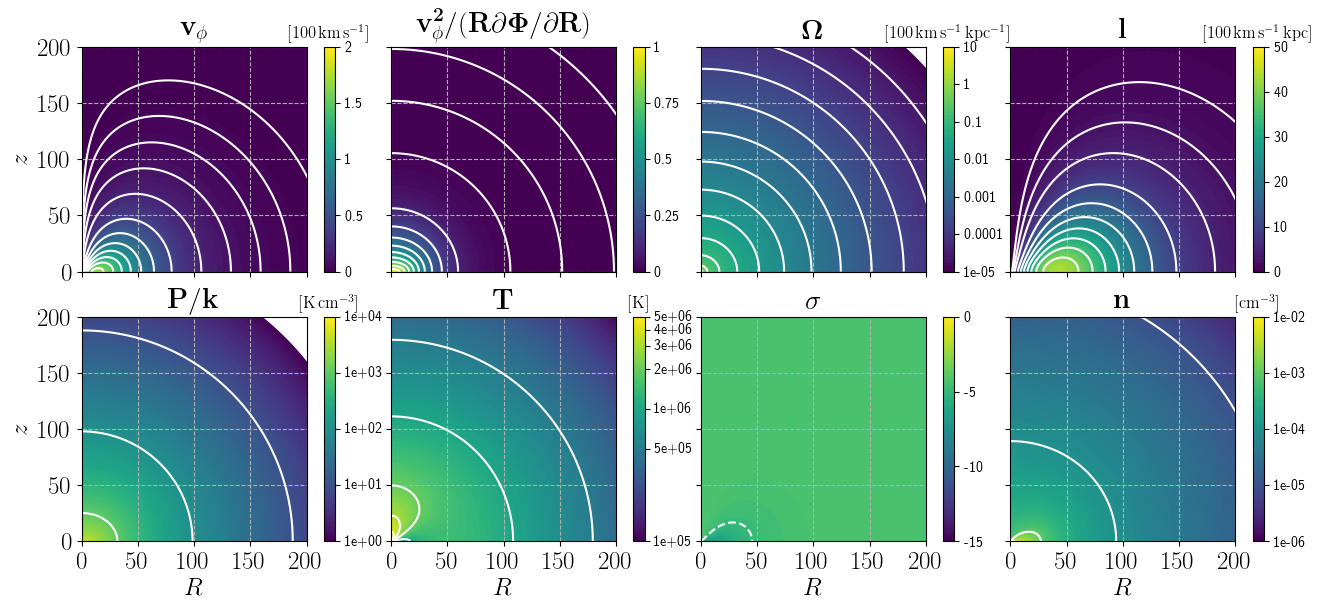}
\caption{Model 6. For $P/k$, $T$ and $n$, the white contours coincide with labels in the colorbars. The white dashed contours for $\sigma$ are at $\{-5,-4,-3\}$.}
\label{fig:m6}
\end{figure*}

\begin{figure*}
\centering
\includegraphics[width=1.0\textwidth]{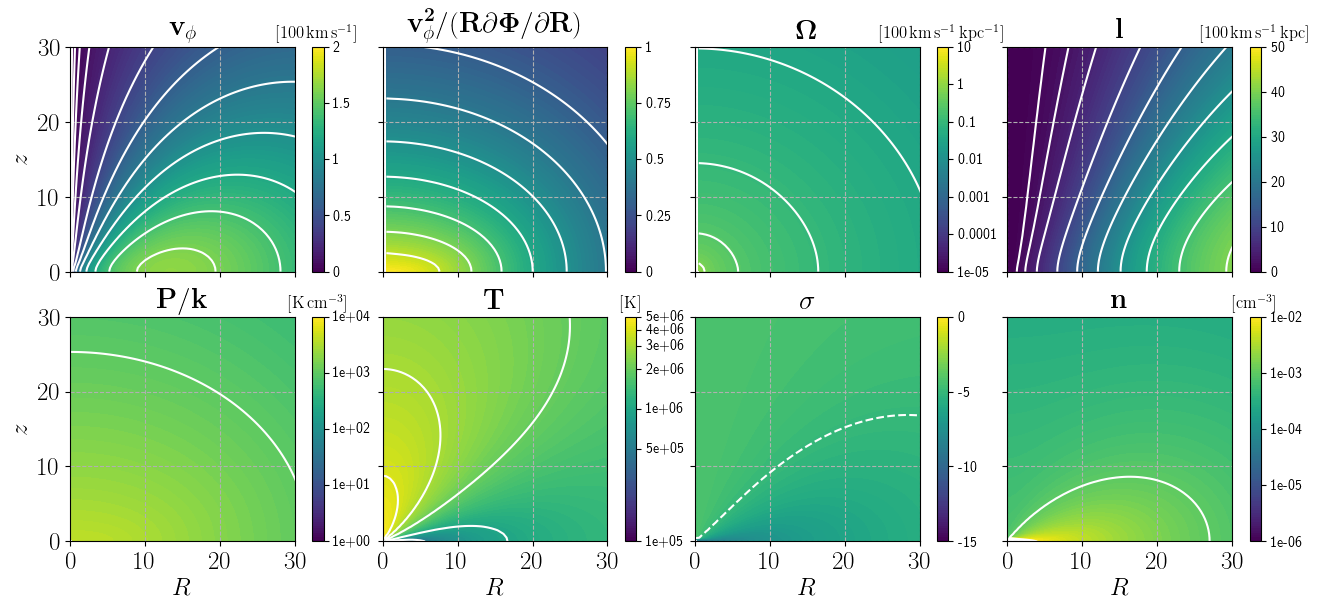}
\caption{Model 6. Zoom in the innermost 30 $\kpc$ of Fig. \ref{fig:m6}.}
\label{fig:m6zoom}
\end{figure*}

This model has $P_{\rm axis}$ as in model 5, thus it does not suffer from the DM problem (Fig. \ref{fig:DMall}), and has $q_{\rm axis}$ as model 3, thus it does not suffer from the AMD problem (Fig. \ref{fig:AMD}). The result is shown in Figs. \ref{fig:m6}, \ref{fig:m6zoom}. This model is therefore consistent with (i) DM of pulsars with known reliable distances; (ii) the densities estimates in Fig. \ref{fig:panels}; (iii) estimates of the rotation velocity close to the plane which show it rotates roughly $80\kms$ slower than the disc \citep{Marinacci+2011,HodgesKluck+2016}; (iv) the roughly exponential AMD profile found in cosmological simulations \citep{SharmaSteinmetz2005}. 

An interesting feature of this model is that it has higher temperature lobes centred on the $z$ axis and close to the Galactic plane, reminiscent of the Fermi bubbles \citep{BHCohen2003,Su+2010}. By looking at X-ray absorption lines \cite{MillerBregman2013} find that, while in most directions their data shows little or no O{\sc VIII} absorption, in the direction of the Fermi bubbles ($l=338.18\degree$, $b=-26.71\degree$) there is an enhancement of O{\sc VIII}. Since O{\sc VIII} is visible only at very high temperature ($T \simeq 4 \times 10^{6} \ \, \rm K$, see for example \citealt{sd93}), this suggests that the temperature of the corona is significantly higher in the direction of the Fermi bubbles. Indeed, by analysing X-ray emission, \cite{Kataoka+2013,Kataoka+2015} and \cite{MillerBregman2016} find that in the direction of the Fermi Bubbles the temperature rises from $T \sim 2\times 10^6 \ \rm K$ to $T \sim 4\times 10^6 \ \rm K$. Our models would be consistent with these expectations, and it would be interesting to explore what dynamical effects these high temperature lobes have once the models are allowed to evolve in time under the presence of a slow cooling and/or thermal conduction. We are not claiming that the Fermi bubbles are a consequence of our model, although we cannot exclude that the corona plays a dynamical role in producing an outflow \citep[e.g.][]{Waxman1978}. However, we note that a rotating halo does favour an outflow compared to a spherical halo, because it has lower density in the directions above and below the Galactic plane than within the plane (see also the models of \citealt{Pezzulli+2017}), thus effectively clearing the way for an outflow.

This model is to a high degree isentropic (see Figs. \ref{fig:m6}, \ref{fig:m6zoom}). This is because we have chosen $\Gamma=5/3$. However, we have chosen this value mainly for simplicity. A model with qualitatively similar characteristics but much farther from being isentropic can be obtained taking for example $\Gamma=1.4$. Thus, we are not ruling out models with substantial entropy gradients.

The spin parameter of all the models in Table \ref{table:1} are in the range $\lambda=0.02\mhyphen 0.04$. These are typical values for dark matter haloes found in simulations \citep[e.g.][]{Bullock+2001,SharmaSteinmetz2005}. However, by analysing a range of simulated galaxies from the {\sc EAGLE} simulations, \cite{Oppenheimer2018} recently found that typical spin parameters of coronae are 2-3 times higher than dark matter spin parameters (see also \citealt{Danovich+2015,Teklu+2015}). Thus it may be worth in the future to explore coronal models with higher spin parameters. This is probably best done using a flattened external potential, which would better represents a rotating dark matter halo than the spherical potential adopted in this paper.

To make progress and construct a more accurate model of the MW, we need to fit parametric models to X-ray surface brightnesses and spectra observations. This requires special care, for example in carefully subtracting contributions due to the Local Bubble \citep{Sanders+1977,CoxReynolds1987}, to the interaction between the Solar wind and interstellar neutrals \citep[e.g.][]{Cravens2000,Liu+2017} and to other Galactic and extragalactic sources, which is out of the scope of the present paper.

\section{Conclusions and outlook} \label{sec:conclusion}

We have presented a simple method to construct general analytic equilibrium baroclinic models of galactic coronae with realistic rotations. We have considered the particular class of models whose equipressure surfaces are ellipses. These models are completely determined by the two functions $P_{\rm axis}$ and $q_{\rm axis}$ which specify the pressure and axis ratio along the axis $(R=0,z)$. This class of models is quite broad and can produce vastly different rotational, density and temperature profiles. Thus it is likely that the sparse observations available can be fitted by a model of this type. Importantly, the models are computationally cheap and suited to be used in fitting algorithms and/or large parameter scans.

As an illustration of the models, we have taken the first step towards fitting dynamical models to the corona of the Galaxy. By a trial and error process, we have constructed models which are compatible with an increasing number of constraints.
We have finally presented a model (number 6) which is consistent with (i) DM of pulsars with known reliable distances; (ii) the densities estimates listed in Fig. \ref{fig:panels}; (iii) the estimates of rotation velocity close to the plane being $80\kms$ slower than those of the disc \citep{Marinacci+2011,HodgesKluck+2016}; (iv) the roughly exponential Angular Momentum Distribution (AMD) found in cosmological simulation \citep[e.g.][]{SharmaSteinmetz2005}.

The next steps are fitting increasingly complicate equilibrium models in order to exploit all the observational data available, in particular X-ray observations \citep{MillerBregman2013,MillerBregman2015,HodgesKluck+2016}. This will unveil the structure of the corona in our own an other galaxies. The subsequent step will be to understand how these models evolve under the presence of a slow cooling and/or thermal conduction, and thus their connection of the problem of accretion onto the Galaxy \citep[][]{PezzulliFraternali2016} and how the gas reservoir necessary to maintain star formation is replenished \citep[e.g.][]{KlessenGlover2016}.

\section*{Acknowledgements}

The authors thank Jeremy Bailin, Filippo Fraternali, Mordecai Mac Low, John Magorrian, Antonino Marasco, Steve Shore, Robin Tress and Freeke van de Voort for illuminating comments and discussions. MCS and RSK 
acknowledge support from the Deutsche Forschungsgemeinschaft via the Collaborative Research Centre (SFB 881) ``The Milky Way System'' (sub-projects B1, B2, and B8) and the Priority Program SPP 1573 ``Physics of the Interstellar Medium'' (grant numbers KL 1358/18.1, KL 1358/19.2, and GL 668/2-1). ES acknowledges support from the Israeli Science Foundation under Grant No. 719/14. GP acknowledges support from the Swiss National Science Foundation grant PP00P2\_163824. RSK furthermore thanks the European Research Council for funding in the ERC Advanced Grant STARLIGHT (project number 339177).

%%%%%%%%%%%%%%%%%%%%%%%%%%%%%%%%%%%%%%%%%
\def\aap{A\&A}\def\aj{AJ}\def\apj{ApJ}\def\mnras{MNRAS}\def\araa{ARA\&A}\def\aapr{Astronomy \&
 Astrophysics Review}\def\apjs{ApJS}\def\apjl{ApJ}\def\pasj{PASJ}\def\nat{Nature}\def\prd{Phys. Rev. D}
\def\ssr{Space Sci. Rev.}\def\pasp{PASP}\def\pasa{Publications of the Astronomical Society of Australia}
\bibliographystyle{mn2e}
\bibliography{bibliography}

\end{document}